# Better results with homogeneous biological macromolecules

Bernard LORBER*

*Architecture et Réactivité de l'ARN, UPR9002, Institut de Biologie Moléculaire et Cellulaire du CNRS & Plateforme de Biophysique Esplanade ARBRE-MOBIEU, 15 rue René Descartes, 67084 Strasbourg, France

Correspondence email : B.Lorber@ibmc-cnrs.unistra.fr

**Abstract**: Pure and homogeneous biological macromolecules (i.e. proteins, nucleic acids, protein-protein or protein-nucleic acid complexes, and functional assemblies such as ribosomes and viruses) are the key for consistent and reliable biochemical and biophysical measurements, as well as for reproducible crystallizations, best crystal diffraction properties, and exploitable electron microscopy images.

**Highlights**:

- Pure and homogeneous macromolecules are the key for the best experimental results

- They warrant the consistency and the reliability of biochemical and biophysical data

- They give more reproducible crystallography and electron microscopy results as well

## 1. Introduction

The results of biochemical and biophysical analyses performed on pure biological macromolecules (e.g. proteins, nucleic acids, high affinity complexes and functional assemblies like ribosomes or viruses) are in principle repeatable. This includes macromolecular concentrations, hydrodynamic properties such as sedimentation velocities and diffusion coefficients, binding parameters such as stoichiometries and affinities, enzymatic specific activities, thermodynamic parameters such as free energies, enthalpies and entropies, or any other property. The same holds for optimized crystallization conditions that produce every time approximately the same number of crystals with comparable sizes, shapes and the least number of growth defects. These crystals exhibit Bragg reflections that are round diffraction spots on the detector, their diffraction limit is always high and their mosaicity low. Electron microscopy images of macromolecules spread over holey carbon films for structure determination do not show aggregated macromolecules or particles. However, it sometimes happens that repetition of the analyses with other macromolecular batches gives results that deviate too much from the average to be attributable to experimental error. In the worst case, the results are at the origin of the discrepancies between published data. The lack of consistent or reproducibility is frequently a consequence of sample heterogeneity. This short note examines the possible causes of heterogeneity and gives practical tips to help beginners improve the quality of their samples and the analysis and interpretation of their experimental data, too.

## 2. Purity vs homogeneity

The purity of proteins and of nucleic acids is very often decisive for their crystallization and for rigorous structural and functional studies as well (see e.g. [1,2]). By definition, pure biological macromolecules and pure particle populations do not contain any contaminant. In other words, all molecules have the same amino acid or nucleotide sequence and all particles belong to a single species. To qualify as pure, samples must pass three tests. Protein and nucleic acids must migrate as a single population on a denaturing electrophoresis gel (containing respectively an ionic detergent or a high concentration of urea). They must have the same hydrodynamic properties in size-exclusion chromatography and the same mass/charge ratio in mass spectrometry. The latter analysis detects mixtures of close sequences that could result from an uncontrolled hydrolysis by a protease or a nuclease, respectively, and minor differences in post-translational or post-transcriptional modifications.

The above three criteria are however not always sufficient (see e.g. [3]). For instance, over 90% of the pure samples analyzed each year in our facility are heterogeneous, i.e. composed of more than one population of macromolecules or one type of particles. Some are blends of free and bound molecules when the number of partner molecule is insufficient. Others contain aggregates that are clusters of randomly associated macromolecules. The size of these aggregates ranges from that of dimers up to that of particles composed of thousands of units. Errors in the folding of polypeptide chains during in vivo biosynthesis and incomplete folding of polynucleotide chains in the test tube may lead to the appearance of ill-defined assemblies. The decrease of macromolecular solubility after a change in concentration, pH or temperature can trigger aggregation. Most of the time, stirring, freezing/thawing and freeze-drying denature the macromolecules and cause aggregation.

Under unfavorable conditions, from less than 1% to up to 100% of the macromolecules in a freshly purified sample can be randomly associated with each other via electrostatic, hydrophobic and/or covalent interactions. Alike misfolded, damaged and inactive macromolecules, aggregates lead to an overestimation of the concentration of functional and active macromolecules if their absorbance spectra overlap. They interfere with the process of crystallization, as do defective molecules. Alike the latter they behave as impurities but their negative effect may be more deleterious. They shift the solubility/crystallization phase diagrams towards precipitation, favor heterogeneous crystal nucleation, and generate, when they incorporate in the lattice, growth defects that end in poor diffraction properties (e.g. low resolution and high mosaicity).

Heterogeneity varies from batch to batch because a human experimenter seldom performs all steps of a purification protocol in a strictly identical way, even if he or she works with the same amount of cells and uses the same automated chromatographic steps. Small variations in mean cell age, protein expression level, dialysis time, protein yield and final sample volume and concentration, suffice to make a detectable difference. Nucleic acids synthesized in a standardized manner may contain variable amounts of incorrectly folded molecules, of abortive synthesis products or counter-ions. Human factors, such as a misinterpretation of a written procedure or an insufficient experimental rigor, enhance the deviation between the compositions of the final products of a same preparation protocol. Hence, and since aggregation is rarely reversible, the preparation of homogeneous macromolecules is the best way to eliminate the counterproductive effects due to variable sample heterogeneity.

## 3. Diagnostic tools

Electrophoresis and isoelectric focusing, high-resolution size exclusion chromatography, mass spectrometry and dynamic light scattering analysis, are suitable for investigating macromolecular heterogeneity under native conditions. Dynamic light scattering measurements take less than a minute and require only microliters of a solution at 1 mg/mL of a protein with $M_r$~100,000. They use non-invasive visible laser light to record the fluctuations of the intensity scattered by the particles subjected to Brownian motion. Then a correlator generates an autocorrelation function that an algorithm converts into a particle size distribution. Single populations of particles give monomodal distributions and those that are monodisperse give narrow ones (Fig. 1). The comparison of the size distributions by intensity and by mass (or volume) informs about the proportion of matter contained in the aggregates.

## 4. Macromolecule preparation and handling

The goal of any molecular and structural biology project is to collect information about how macromolecules are and function in vivo. This imposes to study macromolecules that are structurally and functionally as close as possible to the natural state. To do so, it is necessary to eliminate all sources of heterogeneity due to the artificial conditions that destabilize and insolubilize them. Table 1 provides some tips on how to proceed.

It may be useful to map the temperature and pH ranges in which the macromolecule is stable before reconsidering the steps of the purification protocol. Proteins are stable only between the temperatures at which "cold denaturation" and "heat denaturation" occur. This does not mean that enzymes do not need some time to become fully active after a temperature jump within this range. Nucleic acids are usually more soluble than proteins but their solubility has also limits. A long nucleic acid unfolded at 90°C and refolded on ice, may be insoluble at 4°C but completely soluble at 20°C. The pH is the other parameter to keep constant. The nature and the concentration of the buffering substance define the buffering strength. The preference is for buffers whose $pK_a$ does not vary much with temperature. Protein solubility as a function of pH follows an inverted bell-shaped curve with a minimum at the isoelectric point and higher values below and above le latter. A pH variation by a single unit can have a dramatic effect on solubility.

Ligands or partner molecules may enhance the stability of macromolecules. Apo-enzymes are usually less stable than holo-enzymes. Substrates, analogues of the transition state, or inhibitors can stabilize the conformation of an enzyme. A small coenzyme can extend its half-life from a few minutes to more than a week. Osmolytes like glycerol, as well as salts and short alcohols occupy free binding sites and so increase the solubility of proteins. Non-ionic detergents (at concentrations below their critical micellization concentration) bind to hydrophobic patches located at the surface of proteins that are neither anchored inside nor bound to membranes.

Mobile and disordered regions are frequently deleterious to the formation of packing contacts during crystallization. Post-translational modifications such as glycosylation may introduce more or less heterogeneity and it may be easier to study unmodified proteins. Crystallization phase diagrams of compact macromolecules help find crystallization conditions producing biggest crystals with fewest defects (see e.g. [4]). Crystals grown in a viscoelastic gel may be of superior quality in terms of volume, internal order and stability, with regard to those grown in solution. The polymer chain network of a low-concentration agarose gel does not prevent the formation of a highly ordered crystalline lattice. On the contrary, it has the advantage to stabilize the latter by reduce the effect of osmotic shocks accompanying the soaking with ligands (see e.g. [5]) (**Fig. 2**).

Finally, an ultracentrifugation (of e.g. 1 hour at 100,000 x $g$) suffices to sediment aggregates with dimensions above 100 nm. Analyses based on dynamic light scattering are useful to compare fresh versus older samples and to monitor the solubility and the stability of macromolecules. For best measurements, a centrifugation (10 min at 500 x $g$ in a tabletop centrifuge) of the samples transferred plastic or quartz cuvettes removes microscopic air bubbles and settles undesired dust particles.

## 2. Conclusion

The preparation of homogeneous macromolecules is a cost-effective investment because the latter yield more reproducible and more reliable data in a shorter time. Dynamic light scattering is a simple and powerful method to verify macromolecular homogeneity before undertaking useless and time-consuming investigations. It is also convenient for determining particle size and form factor, detecting weak interactions between macromolecules and monitoring the formation of functional complexes. Its results can be refined using multi-angle light scattering or small-angle X-ray scattering analyses on macromolecules separated by size-exclusion chromatography.

Guidelines for the optimization of data quality are available on the ARBRE-MOBIEU internet page [6] with the quality controls recommended to establish the purity, homogeneity and relative amount of functionally active recombinant proteins, as well as their consistency from batch-to-batch and their long-term stability. A similar approach is desirable for nucleic acids and larger biological entities.

**Acknowledgements**

The author acknowledges funding from the Université de Strasbourg dedicated to the purchase of light scattering instruments.

B. Lorber has compiled the tips listed in Table 1, performed all light scattering analyses reported in Figure 1, grown the crystals displayed in Figure 2, and written the manuscript.

**Table 1**  Hints for better results

_________________________________________________________________________

**General**

- Filtration of buffer solutions on 0.2 μm membranes removes foreign particles
- Fresh macromolecules are more homogeneous than older ones
- Centrifugation of macromolecules removes aggregates
- Size exclusion chromatography removes unwanted molecules
- Complexes crystallize more readily than their components
- A phase diagram helps find best crystallization conditions
- Crystals in agarose gel are three-dimensional and of superior quality
- Agarose gel facilitates crystal soaking with ligands
- Some microfluidic chips are suitable for in situ diffraction analysis

**Proteins**

- Few proteins feel well in pure water
- Protease inhibitors prevent uncontrolled proteolysis
- Stirring, freezing/thawing and freeze-drying have deleterious effects
- Stability range is limited to T(cold denaturation) < T < T(heat denaturation)
- Buffers with lowest $\Delta pK_a$ keep pH constant across a wider T range
- Most assemblies containing proteins behave as proteins
- Reducing agents prevent oxidation
- Protein solubility is lowest at the isoelectric point
- Solubility varies with solvent composition and temperature
- Low concentrations of salt or alcohol may increase solubility
- Low concentrations of non-ionic detergent increase solubility
- Osmolytes such as glycerol may increase solubility
- Ligands and partners molecules increase stability
- Mobile and disordered domains hamper crystallization
- Large post-translational modifications may be heterogeneous
- Thermostability favors crystallizability

**Nucleic acids**

- Nuclease inhibitors prevent unwanted cleavage
- Some ions promote chemical hydrolysis
- Divalent ions and polyamines stabilize the structure
- Minor changes in nucleotide sequence may increase stability
- Post-transcriptional modifications increase molecular rigidity

**Figure 1:** Macromolecular heterogeneity analyzed by dynamic light scattering.

Percentage of the total scattered intensity as a function of particle size before (red/grey curve) and after (black curve) four treatments. (A) Low-speed centrifugation of a ribosome (10 min, 500 x *g*); (B) Ultracentrifugation of a plant protein (1 hr, 100,000 x *g*) ; (C) One cycle of freezing/thawing of a small bacterial protein; (D) Saturation of a virus capsid by a small protein. In every case, the treated sample is homogeneous.

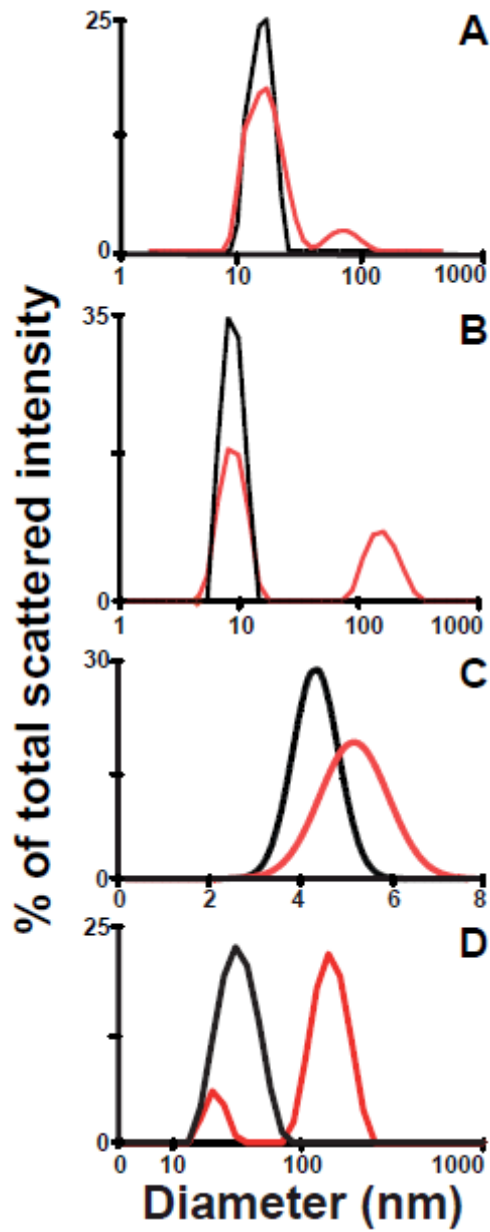

**Figure 2:** Crystals of a human enzyme grown in an agarose gel. Length is 0.1 - 0.3 mm.

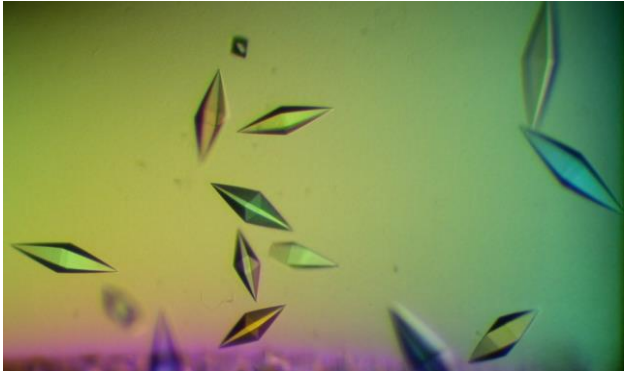